\documentclass{emulateapj}
\usepackage{apjfonts}
\usepackage{graphicx}
\usepackage{epsf}
\usepackage{epsfig, natbib, graphicx, color}

\newcommand{\mbar}{{\bar m}}
\newcommand{\avg}[1]{\langle#1\rangle}
\newcommand{\Mobs}{M_{\rm obs}}
\newcommand{\Mf}{M_{\rm f}}
\newcommand{\Nf}{N_{\rm f}}
\newcommand{\dndlnM}{\frac{dn}{d\ln M}}

\newcommand{\sigmaprior}{\sigma_{\rm prior}}
\newcommand{\hiMsun}{{h}^{-1}M_\odot}
\newcommand{\lnMo}{\ln M_{\rm obs}}
\newcommand{\lnMf}{\ln M_{\rm f}}
\newcommand{\lnM}{\ln M}
\newcommand{\sigmao}{\sigma_{\rm obs}}
\newcommand{\sigmaf}{\sigma_{\rm f}}
\newcommand{\Var}{\mbox{Var}}

\newcommand{\N}{{\bf N}}

\begin{document}

\journalinfo{The Astrophysical Journal, 713:1207-1218, 2010 April 20}
\submitted{Received 2009 July 16; accepted 2010 March 8; published
2010 March 31}
\shortauthors{WU, ROZO, \& WECHSLER}
\shorttitle{CLUSTER FOLLOW-UPS AND DARK ENERGY CONSTRAINTS}

\title{Annealing a Follow-up Program: Improvement of the Dark Energy
Figure of Merit for Optical Galaxy Cluster Surveys}

\author{Hao-Yi Wu\altaffilmark{1}, 
Eduardo Rozo\altaffilmark{2, 3, 4},
Risa H. Wechsler\altaffilmark{1}}

\altaffiltext{1}{Kavli Institute for Particle Astrophysics and
Cosmology, Physics Department, and SLAC National Accelerator
Laboratory, Stanford University, Stanford, CA 94305, USA;
hywu@stanford.edu, rwechsler@stanford.edu}

\altaffiltext{2}{The Center for Cosmology and Astro-Particle Physics,
The Ohio State University, Columbus, OH 43210, USA}

\altaffiltext{3}{The Kavli Institute for Cosmological Physics,
Chicago, Illinois 60637, USA; erozo@kicp.uchicago.edu }

\altaffiltext{4}{Einstein Fellow}

\begin{abstract}

The precision of cosmological parameters derived from galaxy cluster
surveys is limited by uncertainty in relating observable signals to
cluster mass.  We demonstrate that a small mass-calibration follow-up
program can significantly reduce this uncertainty and improve
parameter constraints, particularly when the follow-up targets are
judiciously chosen.  To this end, we apply a simulated annealing
algorithm to maximize the dark energy information at fixed
observational cost, and find that optimal follow-up strategies can
reduce the observational cost required to achieve a specified
precision by up to an order of magnitude.  Considering clusters
selected from optical imaging in the Dark Energy Survey, we find that
approximately 200 low-redshift X-ray clusters or massive
Sunyaev--Zel'dovich clusters can improve the dark energy figure of
merit by $50\%$, provided that the follow-up mass measurements involve
no systematic error.  In practice, the actual improvement depends on
(1) the uncertainty in the systematic error in follow-up mass
measurements, which needs to be controlled at the $5\%$ level to avoid
severe degradation of the results; and (2) the scatter in the optical
richness--mass distribution, which needs to be made as tight as
possible to improve the efficacy of follow-up observations.

\end{abstract}

\section{Introduction}

The dynamical properties of dark energy can be constrained with two
phenomena. The first is the expansion of the universe: dark energy has
dominated the energy density of the universe for the past 4 billion
years and has accelerated its expansion.  The second is the growth of
structure; since dark energy counteracts gravitational attraction, it
slows the growth of structure.  Galaxy cluster surveys explore both
phenomena at the same time: the abundance and the correlation function
of galaxy clusters depend on expansion history and structure growth,
thus providing powerful probes of dark energy.  Given the statistical
power of ongoing and future surveys, galaxy clusters have become an
indispensable probe of dark energy (e.g.\@
\citealt{Wang98,Haiman01,Holder01,Levine02,Hu03, Rozo07a,Rozo07b,
Rozo09}, and references therein).

Four cluster detection methods have been well established: the
intracluster hot gas can be identified via X-ray
\citep[e.g.,][]{Ebeling98, Ebeling00,Ebeling01, Vikhlinin98,
Bohringer04} or Sunyaev--Zel'dovich (SZ) effects \citep[e.g.,][see
also \citealt{Carlstrom02}]{Staniszewski08,Hincks09}; the mass
concentrations can be identified using weak lensing shear
\citep[e.g.,][]{Wittman01,Wittman06}; or the galaxies in clusters can
be identified in optical or infrared surveys
\citep[e.g.,][]{Postman96, Koester07, Eisenhardt08}.  Large cluster
surveys using each method are ongoing or forthcoming, and cosmological
parameter constraints from cluster surveys have recently become
competitive with other dark energy probes
\citep[e.g.,][]{Mantz08,Henry09,Rozo09,Vikhlinin09}.

The key issue for extracting cosmological information from clusters is
the fidelity of the mass tracer.  In a survey, the mass tracer can be
self-calibrated by combining the information from counts and sample
variance \citep{LimaHu04}.  In addition, if a sample or sub-sample of
clusters is observed using multiple methods, the cluster mass can be
further calibrated
\citep[e.g.,][]{Majumdar03,Majumdar04,Cunha08,Cunha09}. Determining
the most effective approach to improve the constraining power of
clusters using multiple mass tracers is particularly timely, as
multi-wavelength observations will soon become available for large
cluster samples.

In this work, we focus on follow-up observations for optical cluster
surveys.  We are particularly interested in how dark energy
constraints from these surveys can be improved when a sub-sample of
the clusters has better mass measurements from other methods, e.g.,
X-ray or SZ.  Our goal is to characterize how follow-up observations
should be designed and what precision is required in order to maximize
dark energy information.

Optical surveys identify massive clusters as agglomerations of
galaxies.  Since the physics of galaxy formation is much more
complicated than the physics of hot intracluster gas, optical mass
tracers are not as well understood as X-ray or SZ mass tracers.
Nevertheless, the optical richness--mass distribution can be
empirically determined, and precise cosmological parameters have been
derived from optically selected cluster samples
\citep[e.g.,][]{Gladders07,Rozo09}.  In the near future, optical
surveys such as the Dark Energy Survey (DES\footnote{\tt
http://www.darkenergysurvey.org}), the Panoramic Survey Telescope $\&$
Rapid Response System\footnote{\tt http://pan-starrs.ifa.hawaii.edu},
and the Large Synoptic Survey Telescope\footnote{\tt
http://www.lsst.org} will be able to identify substantially larger and
higher-redshift cluster samples, which will significantly improve our
knowledge of dark energy equation of state $w$.

In this work, we assume the statistical power and parameters relevant
to DES, apply the self-calibration method proposed by \cite{LimaHu04}
to calculate our fiducial cosmological parameter constraints and
explore how follow-up observations will improve dark energy
constraints. Throughout, we quantify dark energy constraints with the
figure of merit (FoM) proposed in the Report of the Dark Energy Task
Force \citep{Albrecht06}:
\begin{equation}
{\rm FoM}=1/\sqrt{{\rm det}\ {\rm Cov}(w_0,w_a)}=[\sigma(w_a)\sigma(w_p)]^{-1} \ ,
\end{equation}
where $w=w_0+(1-a) w_a$ and $w_p$ is calculated at the pivot redshift
for which $w$ is best constrained. The current value of the FoM from
WMAP5, SNe, and BAO is 8.326 \citep{WangY08}.  As a reference, the
DETF report predicts that the FoM from Stage III cluster surveys (of
which DES is an example) will range from 6.11 to 35.21, depending on
the prior on the observable--mass distribution.

Assuming that the observable--mass distribution follows power laws in
both mass and redshift, we find that the FoM expected from a DES-like
survey is 15.5 using a self-calibrated analysis.  We apply a simulated
annealing algorithm to design follow-up strategies that maximize the
FoM, starting with the limiting case in which follow-up mass
measurements have infinite precision. We then study how the FoM
improvements depend on real-world complications, finding that the
efficacy of follow-up observations is likely to be limited by the
systematic error in follow-up mass measurements. We finally consider
observational issues and design different cost-effective follow-up
strategies for X-ray and SZ.

\cite{Majumdar03,Majumdar04} have previously investigated how
follow-up mass measurements of a fraction of the X-ray or SZ selected
cluster sample can constrain the cluster evolution and improve dark
energy constraints.  Our major improvement upon these studies is the
optimization of follow-up strategies.  We also include counts-in-cells
and mass binning, and explore how scatter and possible systematic
error of mass tracers can affect the efficacy of follow-up
observations.

\cite{Cunha08} has recently studied the joint analysis of overlapping
optical and SZ cluster surveys.  By studying the cluster abundances in
both surveys, the observable--mass distribution can be
cross-calibrated.  In contrast to this study, we concentrate on one
survey and its follow-up observations. That is, our method does not
require another complete survey but instead focuses on a small and
optimized follow-up program.

This paper is organized as follows.  We briefly review dark energy
constraints from clusters and the self-calibration analysis in
Section~\ref{sec:self-calibration} and describe our survey and model
assumptions in Section~\ref{sec:assumptions}.  We present the modeling
for follow-up observations in Section~\ref{sec:follow-up} and the
optimization procedure in Section~\ref{sec:opt}. Real-world
complications of follow-up observations are explored in
Section~\ref{sec:complications}.  Optimizations considering
observational issues specific to X-ray and SZ clusters are carried out
in Section~\ref{sec:observability}. We discuss other relevant studies
in Section~\ref{sec:discussion} and conclude in
Section~\ref{sec:summary}.


\section{Dark Energy Constraints from Galaxy Clusters}
\subsection{Self-calibration: A Review}
\label{sec:self-calibration}

In this section, we briefly review the self-calibration formalism
developed by \cite{LimaHu04}.  For detailed discussions, we refer the
reader to \citet{LimaHu04,LimaHu05,LimaHu07}, \citet{HuCohn06}, and
\citet{WuHY08}.

In a galaxy cluster survey, one studies how cluster counts depend on
some cluster mass proxy to infer the dark matter halo mass function,
which provides constraints on the properties of dark energy.  To infer
the mass function from data, one needs to relate the observable
properties of clusters to the halo mass; thus, the uncertainty in this
observable--mass distribution limits the constraining power of
clusters.  The observable--mass distribution can be self-calibrated
using a counts-in-cells analysis, in which the survey volume is
divided into small cells and the halo bias can be calculated from the
sample variance among the cells.  By measuring both counts and sample
variance, the mass function and the halo bias are fit simultaneously,
the observable--mass distribution can be self-calibrated, and the dark
energy constraints can be improved.

Let us consider a cell of volume $V_{\rm c}$ in a narrow redshift
range in a survey.  We denote the cluster mass proxy by $\Mobs$ and
further bin our cluster sample in $\Mobs$ with a binning function
$\phi(\ln\Mobs)$, which equals unity inside the bin and zero
elsewhere.  In this bin, the mean cluster counts ($\bar m$) and the
mean cluster bias ($\bar b$) can be calculated from the mass function
$dn/d\ln M$, the halo bias $ b(M)$, and the observable--mass
distribution $P(\ln\Mobs|\ln M)$:
\begin{eqnarray}
\bar m &=& V_{\rm c} \int d\ln M\ \dndlnM \avg{\phi|\ln M} \ , \label{eq:counts}\\
\bar  b &=& \frac{V_{\rm c}}{\bar m}\int d\ln M\ \dndlnM  b(M)  \avg{\phi|\ln M} \label{eq:bias}\ ,
\end{eqnarray}
where 
\begin{equation}
\avg{\phi|\ln M} = \int d\ln\Mobs\ P(\ln\Mobs|\ln M) \phi(\ln\Mobs) \ . 
\end{equation} 
We assume the redshift bin is thin enough and simply use the
function values at the middle redshift instead of averaging over the
redshift bin.

The mean cluster bias $\bar b$ determines the number fluctuations
among cells due to the large-scale clustering.  If cell $i$ has
cluster counts $m_{i}$ and bias $\bar b_{i}$, the corresponding sample
variance among the cells is given by
\begin{eqnarray}
S_{ij}&=& \avg{(m_{i}-\mbar_i) (m_{j}-\mbar_j)} \\
	&=& \mbar_i \mbar_j \bar  b_{i} \bar  b_{j}\sigma_{\rm V_{\rm c}}^2 \ , \nonumber
\end{eqnarray}
where $\sigma_{\rm V_{\rm c}}^2$ is the variance of the dark matter
density fluctuation in volume $V_{\rm c}$.  We assume that the cell
volume is large enough for the covariance between cells to be
negligible.

In a survey, one observes $\bar m_i$ and $S_{ij}$, calculates $\bar
b_i$, and self-calibrates $P(\ln\Mobs|\ln M)$ based on
Equations~\ref{eq:counts} and~\ref{eq:bias}.  We use matrix notations
${\bf \bar m}$ and ${\bf S}$ to indicate the data in multiple mass and
redshift bins and further define ${\bf C} = {\rm diag}({\bf\bar m})
+{\bf S}$.  The constraints on model parameters can be obtained from
the Fisher matrix
\begin{equation}
F_{\alpha \beta}=\bold{\bar m}^{T}_{,\alpha} \bold C^{-1} \bold{\bar m}_{,\beta} + \frac{1}{2} {\rm Tr}[\bold C^{-1} \bold S_{,\alpha}\bold C^{-1} \bold S_{,\beta}] 
\ ,
\label{eq:fisher}
\end{equation}
where the comma followed by a subscript indicates the partial
derivative with respect to a model parameter, and the derivatives are
with respect to cosmological parameters and the parameters
characterizing the observable--mass distribution.  We invert this
Fisher matrix to obtain the covariance matrix and constraints on model
parameters.

The second term in this Fisher matrix characterizes the information
from the sample variance.  Since the sample variance depends on both
cosmology and observable--mass distribution, this ``noise'' in cluster
counts actually provides ``signal.''  As we will see in
Section~\ref{sec:follow-up}, an analogous Fisher matrix is used to
calculate the constraints from follow-up observations.  In those
observations, the variance in follow-up mass measurements will provide
critical information for breaking the degeneracies between model
parameters.


\subsection{Survey and Model Assumptions}
\label{sec:assumptions}

In this work, we consider a DES-like optical survey with a survey area
of $ 5000\deg^2$, and a survey depth such that clusters with redshift
$z<1$ are robustly identified.  For the counts-in-cells analysis, the
survey volume is divided into cells of area $=10\deg^2$ and $\Delta z
=0.1$ \citep[see, e.g.,][]{LimaHu04}.  We assume that the mass
threshold for the survey is $\Mobs=10^{13.7}\ \hiMsun$, and that the
cluster sample is binned by $\Mobs$ into eight bins with logarithmic
bin size $\Delta\log_{10}\Mobs=0.2$.  Note that the mass threshold and
binning are based on the mass proxy $\Mobs$ rather than the true
cluster mass $M$.  We have tested the impact of the binning on our
results and found that this binning is fine enough that we do not lose
information from the data.  Finer binning is also unnecessary because
the scatter between the observable and mass sets an effective mass
resolution.

For the observable--mass distribution $P(\ln\Mobs|\ln M)$, we assume a
Gaussian distribution with mean ($\ln M +\ln M_{\rm bias}$) and
variance $\sigmao^2$.  Both $\ln M_{\rm bias}$ and $\sigmao^2$ are
assumed to vary linearly with $\ln M$ and $\ln (1+z)$ via
\begin{eqnarray}
\ln M_{\rm bias} &=& \ln M_0+\alpha_M\ln (M/M_{\rm pivot})+\alpha_z \ln (1+z) \label{eq:Mbias}\\
\sigmao^2 &=& \sigma_0^2+ \beta_M\ln (M/M_{\rm pivot})+\beta_z \ln (1+z)\ , \label{eq:scatter}
\end{eqnarray}
giving nuisance parameters: ($\ln M_0, \alpha_M, \alpha_z, \sigma_0^2,
\beta_m,$ $\beta_z$).  For our fiducial model, we assume an unbiased
and non-evolving observable--mass distribution with $\sigma_0=0.5$
(all other parameters are set to zero), which is consistent with the
results in \cite{Gladders07} and \cite{Rozo09}.  In addition, $M_{\rm
pivot}$ should be close to the mass scales of interest; we choose
$M_{\rm pivot} = 10^{13.7}\hiMsun$, noting that the precise value of
$M_{\rm pivot}$ does not affect our results.

Throughout this work, we assume the {\em Wilkinson Microwave
Anisotropy Probe Five Year (WMAP5)} cosmological parameters
(\citealt{Komatsu09}; $w_0 = -1,\ w_a =0,\ \Omega_{\rm DE} = 0.726,\
\Omega_{\rm k}=0,\ \Omega_{\rm m}h^2=0.136,\ \Omega_{\rm
b}h^2=0.0227,\ n_{\rm s}=0.960,\ \Delta_\zeta = 4.54\times 10^{-5} $
at $k = 0.05 {\rm Mpc^{-1}}$), and use the Planck-prior Fisher matrix
provided by W.\@ Hu and Z.\@ Ma.  We use the linear matter power
spectrum calculated by CAMB \citep{Lewis00}.  The halo mass is defined
with spherical overdensity of $\Delta = 200$ with respect to the mean
matter density, and we use the updated mass function from
\cite{Tinker08} and the halo bias function from
\cite{Sheth01}\footnote{An updated halo bias function is available in
\cite{Tinker10}; the difference between the two functions does not
significantly affect our results.  We also note that the mass function
and the halo bias function we use are not in a consistent framework;
since we only use their dependence on cosmology, this inconsistency is
not important.  }.  Under these assumptions, we expect that a DES-like
survey will observe approximately $2\times10^5$ clusters in total.


\section{Improving Dark Energy Constraints with Optimal Follow-up
  Strategies}
\label{sec:baseline}
\subsection{Constraints from Follow-up Mass Measurements}
\label{sec:follow-up}

In this section, we calculate the additional constraints from
follow-up observations.  Given the optical cluster sample from a
DES-like survey, we select some clusters from each mass and redshift
bin to follow up---for example, to measure their properties in X-ray
or SZ---and estimate the cluster mass more precisely. These
follow-up mass measurements provide further constraints on the
observable--mass distribution, thus improving the dark energy
constraints of the original survey.  Throughout Section~\ref{sec:baseline},
we assume the follow-up observations provide mass measurements with
infinite precision; the complications of follow-up mass tracers will
be explored in Section~\ref{sec:complications}.

Our goal is to constrain the scaling relation and the scatter of
optical richness.  We note that the term ``scatter'' usually has
different meanings in theoretical and observational contexts.  In the
theoretical model, the scatter of $\ln\Mobs$ at fixed $\ln M$
($\sigmao$ in our notation) is used because the model is based on the
mass function, which is a function of $M$.  In contrast, one
observationally constrains the scatter of $\ln M$ at fixed $\ln\Mobs$
because follow-ups are selected based on their $\Mobs$
\citep[e.g.,][]{Rozo09Richness}.  In general, the theoretical model is based
on $P(\ln\Mobs|\ln M)$, while observations put constraints on $P(\ln
M|\ln\Mobs)$.  Our goal is therefore to convert constraints on the
latter distribution to constraints on the former.

Let us return to the original survey, focusing on the follow-up
observations in a bin specified by $\phi(\ln\Mobs)$ in a narrow
redshift range.  If the follow-up mass $\Mf$ perfectly recovers the
true mass, the mean and variance of $\ln\Mf$ will read
\begin{eqnarray}
\avg{\ln \Mf}&=&E[\ln \Mf|\phi(\ln\Mobs)] \nonumber\\
&\propto& \int d\ln M \ln M \dndlnM \avg{\phi|\ln M} \ ,
\end{eqnarray}
and
\begin{eqnarray}\label{eq:var}
&&V=\Var[\ln \Mf|\phi(\ln\Mobs)] \nonumber\\
&&\propto \int d\ln M (\ln M-\avg{\ln \Mf})^2 \dndlnM \avg{\phi|\ln M} \ .
\end{eqnarray}

The information from a single follow-up mass measurement is given by
an analog of Equation~\ref{eq:fisher}:
\begin{equation}\label{eq:fisher1}
\tilde F_{\alpha\beta}^{(i)}=\avg{\ln\Mf}_{,\alpha}V^{-1}\avg{\ln \Mf}_{,\beta} + \frac{1}{2}V^{-1}V_{,\alpha}V^{-1}V_{,\beta}  \ ,
\end{equation}
where the superscript $(i)$ indicates the mass and redshift bin from
which we select follow-ups.  If we follow up $N_{\rm i}$ clusters in
bin $i$, the Fisher matrix for the whole follow-up program reads
\begin{equation}\label{eq:fisher2}
\tilde F_{\alpha\beta} = \sum_i N_{\rm i} \tilde F_{\alpha\beta}^{(i)}.
\end{equation}
This Fisher matrix is added to the Fisher matrix of counts-in-cells
(Equation~\ref{eq:fisher}) to improve the constraints on nuisance
parameters\footnote{{\it Technical note.} In principle, when we
calculate the Fisher matrix $\tilde F_{\alpha\beta}$, the derivatives
should include cosmological parameters.  However, to correctly model
the cosmological information in follow-up observations, we need to
include the covariance between the follow-up observations and the
original cluster survey.  For simplicity, we ignore this covariance
and only consider the information for nuisance parameters, noting that
the cosmological information is negligible for up to 1000
follow-ups.}.  Note that the second term in Equation~\ref{eq:fisher1}
plays the role of ``noise as signal'' as in the case of
counts-in-cells, characterizing the information included in the mass
variance of follow-ups.


\begin{figure}[t!]
\epsscale{1.2}
\plotone{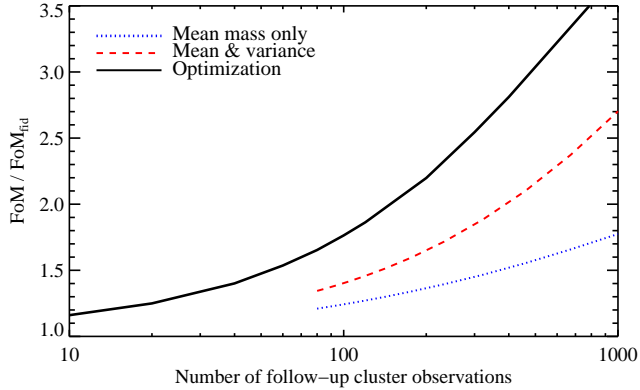}
\caption{ Improvement in the dark energy FoM due to follow-up
observations, relative to the fiducial value ($\rm FoM_{fid}$) for a
DES-like survey.  The black solid curve corresponds to optimal
follow-up strategies that maximize the FoM; the red dashed curve
assumes that the follow-ups are evenly selected from all mass and
redshift bins; the blue dotted curve also assumes even selection,
ignoring the variance of the follow-up mass measurements.  The
follow-ups can significantly improve the FoM, especially when the
target selection is optimized and when the variance of follow-up mass
measurements is included.  }
\label{fig:Nf}
\end{figure}


Figure~\ref{fig:Nf} shows how follow-up observations improve the dark
energy FoM (defined in Section 1).  We calculate the ratio between the
improved FoM and the fiducial FoM from a DES-like survey (without
follow-ups) and demonstrate how this ratio improves as the number of
follow-up mass measurements increases.  For the two lower curves, we
assume that follow-up targets are evenly selected across all bins
(i.e., $N_{\rm i}=\Nf/80$, including eight bins in $\Mobs$ and ten
bins in redshift).  We also limit the number of follow-ups in each bin
by the number of clusters that a DES-like survey is expected to
observe.

The red dashed curve shows the improvement of the FoM using the
information from both the mean and the variance of follow-up mass
measurements (i.e., both terms in Equation~\ref{eq:fisher1}).  As can
be seen, the follow-ups can substantially improve the FoM; for
example, for $100$ follow-up mass measurements, the FoM can be
improved by $40.3\%$.  For comparison, the blue dotted curve shows the
improvement in the FoM using the information only from the mean of
follow-up mass measurements (i.e., only the first term in
Equation~\ref{eq:fisher1}).  As can be seen, lacking the information
from the variance can substantially reduce the effectiveness of
follow-ups. This case may be relevant to stacked cluster samples which
provide the mean but not the variance of cluster mass.  Finally, the
black solid curve corresponding to the most significant improvement is
based on the optimal follow-up strategies, which will be discussed in
the next section.

In addition to evenly selected follow-up targets, we also investigate
the case of selecting a fixed fraction of the cluster sample in each
bin, which is assumed in \cite{Majumdar03}.  This selection gives
slightly lower FoM values than our evenly selected follow-ups; the
reason will become clear in the next section.


\subsection{Optimizing Follow-up Strategies: A Simulated Annealing Approach}
\label{sec:opt}

Instead of selecting the same number of follow-up targets in all bins,
in this section, we explore how we can further improve the FoM by
optimally selecting the follow-ups. We pose the following optimization
problem: for a fixed number of follow-up measurements, $\Nf$, how
should we select follow-up targets to maximize the FoM?

We use a slight variant of the Metropolis-based simulated annealing
algorithm to find the optimal follow-up strategy.  The idea of
simulated annealing comes from a physical phenomenon: when a metal
cools slowly, its atoms will rearrange to achieve the minimum of
internal energy.  This rearrangement sometimes moves toward
higher-energy states due to thermal fluctuations, which allow the
system to overcome energy barriers separating different local minima.
As the system cools, these thermal fluctuations become rare; provided
the cooling process is slow enough, the system will eventually
approach its global energy minimum.

The simulated annealing algorithm is suitable for our problem because
our problem is combinatorial and the total number of configurations is
factorially large. (If we use 80 bins, there are $10^{52}$
configurations for $\Nf=100$ and $10^{121}$ configurations for $\Nf =
1000$.) Following the guidelines in \cite{Press02NR}, we design our
algorithm as follows:

\begin{enumerate}

\item The system configuration is characterized by the number of
follow-up targets in each mass and redshift bin, $\N=(N_1,...,
N_{80})$. (We use eight bins in $\Mobs$ and ten bins in redshift.) We
note that the general configurations do not change with the detail of
binning.

\item The rearrangement of clusters in bins, or the transition from
one configuration to another, is designed as follows: given an initial
configuration, for each ``donor '' bin $i$, we randomly pick a
``receiver'' bin $j$ and transfer $n_{\rm trans}$ clusters to the
receiver bin.  Here $j$ is randomly chosen from all available bins,
and $n_{\rm trans}$ is a random integer between $0$ and $n_{\rm
limit}$.  We choose $n_{\rm limit}$ to be $1\%$ of the total number of
follow-ups $\Nf$; for example, if $\Nf =100$, we transfer $0$ or $1$
cluster at a time.  We also require the number of follow-ups in each
bin to be between zero and the number of clusters expected to observed
in that bin.  After running $i$ through all bins, i.e., letting each
bin play the role of donor once, we reach a new nearby configuration.

\item The objective function is the FoM, which we are trying to
maximize by sampling different configurations.  Applying the idea of
the Metropolis algorithm, we start from the current configuration
($\N_{\rm i}$, with the FoM value $F_{\rm i}$) and sample a nearby
configuration ($\N_{\rm try}$, with the FoM value $F_{\rm try}$). If
$F_{\rm try}>F_{\rm i}$, $\N_{\rm try}$ is accepted at this step; that
is, we set $\N_{\rm i+1} = \N_{\rm try}$.  If $F_{\rm try}<F_{\rm i}$,
$\N_{\rm try}$ is accepted with probability $\exp[(F_{\rm try}-F_{\rm
i})/T]$.  Here $T$ is the ``temperature'' parameter that determines
the probability of moving to a smaller FoM value (analogous to thermal
fluctuations).

\item To design our annealing schedule, we start with a $T$ value that
roughly gives an acceptance rate of $0.2$; this rate empirically
indicates a fair sampling of the configuration space (see, e.g.,
\citealt{Dunkley05} for the case of Markov chain Monte Carlo). After
running $10^4$ iterations with this temperature, we lower the
temperature by $10\%$ and run $4000$ iterations as one step of the
annealing.  We repeat this annealing procedure between 10 and 40 times
(depending on the size of the configuration space) until the
improvement in the FoM is negligible and the system ``entropy'' is low
in the sense that clusters are concentrated in a few bins.

\end {enumerate}

Since our configuration space is factorially large, we reduce its size
to facilitate the sampling.  We start with only 40 bins by doubling
the bin size in $\Mobs$ and determine where the relevant bins are.  We
then return to our original binning and exclude irrelevant bins to
reduce the size of the configuration space.

Regarding the convergence of our algorithm, we note that although the
simulated annealing algorithm will eventually converge to the global
optimum, pursuing this convergence is impractical due to the extremely
large configuration space.  We use a sufficiently high temperature at
the beginning to ensure that the configuration space is fairly
sampled, but we cannot guarantee that the global optimum is found at
the end. However, after testing several different initial
configurations, we find that various local optima are very close to
each other. Each of these optima provides significant improvement when
compared to the evenly selected follow-ups.  We thus expect that our
method provides a good solution to the optimization problem.


\begin{figure}[t!]
\epsscale{1.2}
\plotone{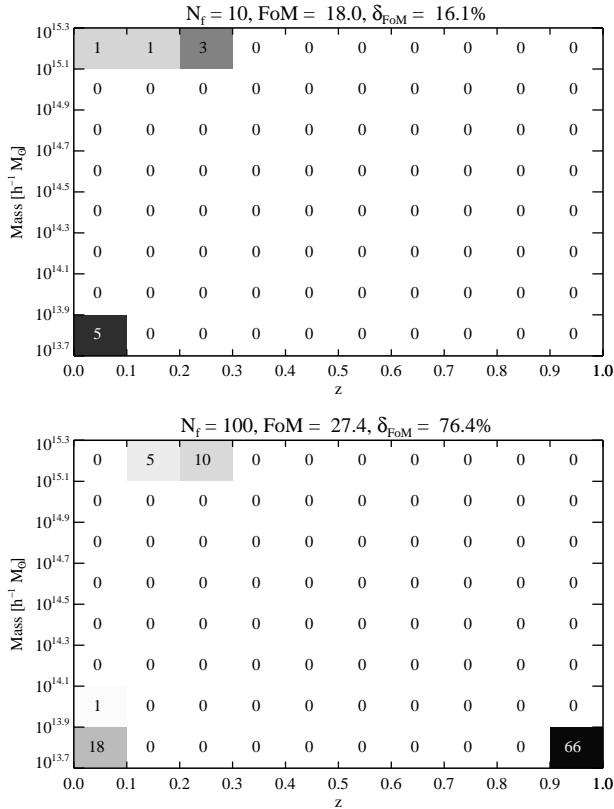}
\caption{ Optimal follow-up strategies that maximize the FoM at a
given number of follow-up observations $\Nf$, calculated with a
simulated annealing algorithm.  Each pixel corresponds to one mass and
redshift bin in the optical cluster survey, and the corresponding
number of follow-ups in this bin is shown.  {\it Upper:} $\Nf = 10$;
{\it lower:} $\Nf = 100$.  As can be seen, low-redshift bins are
highlighted because they provide complementary information to the CMB
prior.  In addition, the follow-up strategies prefer the corners,
because these bins provide the longest lever arms for constraining the
mass and redshift dependence of the observable--mass distribution.  }
\label{fig:opt}
\end{figure}


Figure~\ref{fig:opt} illustrates two examples of our optimization; the
upper panel corresponds to ten follow-ups in total, $\Nf=10$, and the
lower panel corresponds to $\Nf=100$.  For $\Nf=10$, the optimal
configuration focuses on the highest and the lowest mass bins at low
redshift.  This configuration reflects the fact that given the tight
CMB prior, most of the constraining power of a cluster survey lies in
the low-redshift cluster sample, which provides the longest possible
lever arm for constraining the evolution in dark energy.  This optimal
follow-up strategy improves the FoM by $16.1\%$ relative to the
fiducial FoM of the survey.  For $\Nf=100$, high-redshift clusters
start to become important, and the improvement in the FoM is $76.4\%$,
compared to $40.3\%$ for uniform sampling.  We note that in the
highest-mass and lowest-redshift bins, clusters are very rare and we
reach the limit of following up all clusters in these bins.

We can see that the optimal follow-up strategies always select
clusters in the most extreme bins---only the corners of the mass and
redshift bins shown in Figure \ref{fig:opt} are populated. This result
reflects our assumption that the observable--mass distribution depends
on mass and redshift via power laws, which are well constrained by the
extremes.  These strategies also imply that following up a random
fraction of the cluster catalog is inefficient as a way to improve the
FoM, since typical clusters are in the low-mass and mid-to-high
redshift regime.

In practice, following up clusters in the highest-redshift and
lowest-mass bin is impractical, if not impossible.  Consequently, we
explore how superior this bin is relative to its neighbors.  We find
that this bin is not significantly better than other high-redshift and
low-mass bins.  The lowest-redshift bins, on the other hand, are
always the most essential to follow up.  Thus, for designing a
practical follow-up program, one should always prioritize the
lowest-redshift bins and try to extend to the high-redshift and
low-mass regime.  We will further explore the detectability and
observational cost issues to find the most cost-effective strategies
in Section~\ref{sec:observability} .

The improvement in the FoM due to the optimization is shown as a black
solid curve in Figure~\ref{fig:Nf}.  We select various $\Nf$ values
and apply the optimization algorithm to maximize the FoM. Compared to
the case of uniformly sampled follow-ups, the optimization can further
improve the FoM.

For an interesting comparison, we note that \cite{Frieman03} optimized
the supernova survey strategy to minimize the errorbar on $w$.  They
demonstrated different survey strategies to complement different CMB
priors.  We also note that \cite{Parkinson07, Parkinson10} applied
simulated annealing to the design of baryon acoustic oscillation
surveys.


\section{Requirements for the Follow-up Mass Proxy}
\label{sec:complications}
\subsection{Scatter and Covariance of the Mass Proxies}
\label{sec:correlation}

So far we have assumed that follow-up observations can recover the
true mass precisely; i.e., $\Mf=M$.  In reality, $\Mf$ itself is a
mass proxy and has a scatter $\sigmaf$ around the true mass $M$.
Moreover, this scatter may correlate with $\sigmao$, the scatter of
$\Mobs$ around $M$.  Therefore, a proper analysis of the effects of
follow-up observations should include the observable--follow-up--mass
distribution $P(\lnMo,\lnMf|\lnM)$.

We assume that the follow-up mass tracer correlates with the true mass
more tightly than the survey mass tracer does; for example, X-ray and
SZ mass proxies have lower scatter than any known optical mass proxy
\citep[e.g.,][]{Kravtsov06}.  We have assumed $\sigmao = 0.5$
throughout our analysis; we further assume that $\sigmaf = 0.1$ and
that the two mass proxies, $\ln\Mobs$ and $\ln\Mf$, have a correlation
coefficient $\rho$.  No prior knowledge is assumed about $\rho$.

Let us assume that the observable--follow-up--mass distribution
$P(\lnMo,\lnMf|\lnM)$ is a bivariate Gaussian distribution in
$(\lnMo-\lnM)$ and $(\lnMf-\lnM)$ with the covariance matrix
\[
\left(
\begin{array}{cc}
  \sigmao^2& \rho\sigmao\sigmaf     \\
   \rho\sigmao\sigmaf& \sigmaf^2    
\end{array}
\right) .
\]
The mean of $\ln \Mf$ is given by
\begin{eqnarray}
&&E[\lnMf|\phi(\lnMo)]\propto\int d\lnM\frac{dn}{d \lnM}\int d\lnMf\lnMf \nonumber\\
&&\int \lnMo\phi(\lnMo)
P(\lnMo,\lnMf|\lnM)   \ .
\end{eqnarray}
The variance of $\lnMf$ can be calculated similarly.  Since this
variance involves both $\sigmao$ and $\sigmaf$, in the limit
$\sigmaf^2 \ll \sigmao^2$, we expect the variance of $\ln \Mf$ to be
dominated by $\sigmao^2$.  On the other hand, if $\sigmaf$ is larger
(i.e., the follow-up mass measurements have larger intrinsic scatter),
the resulting mass variance of follow-ups will be larger and the FoM
improvement will be less significant.

We find that different values of $\rho$ barely affect the resulting
FoM: positive $\rho$ slightly improves the FoM, while negative $\rho$
slightly degrades the FoM.  When compared to positive $\rho$, negative
$\rho$ widens the mass range of the follow-ups (with larger variance
in $\ln \Mf$) and has slightly less constraining power.  The reason is
that when the correlation is negative, $\sigmao$ and $\sigmaf$ tend to
scatter $\Mobs$ and $\Mf$ to opposite directions, and cluster samples
selected based on $\Mobs$ will have larger variance in $\Mf$.  Since
the differences are small, we only show the case of a zero correlation
for demonstration.


\begin{figure}[t!]
\epsscale{1.2}
\plotone{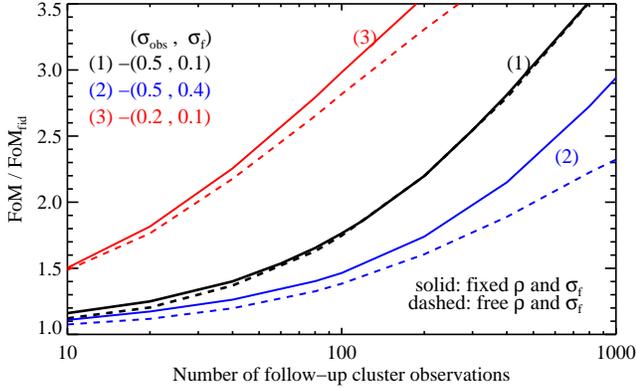}
\caption{ Impact of the scatter and the correlation of mass proxies on
the effectiveness of optimized follow-up observations.  Here we
compare three sets of assumptions about $\sigmao$ and $\sigmaf$; for
each set we show the cases of fixed (solid curves) and marginalized
(dashed curves) $\rho$ and $\sigmaf$.  When $\sigmaf \ll \sigmao$, the
FoM is almost unaffected by marginalization, as shown in (1).  When
$\sigmaf$ is large, the effectiveness of follow-ups is degraded, as
shown in (2).  On the other hand, when the survey has small scatter
(small $\sigmao$), follow-ups can significantly improve the FoM, as
shown in (3).  }
\label{fig:correlation}
\end{figure}


We allow the scatter $\sigmaf$ and the correlation coefficient $\rho$
to be additional free parameters in the Fisher matrix
(Equation~\ref{eq:fisher2}) and study their impact.  In
Figure~\ref{fig:correlation}, we explore how different assumptions of
scatter affect the FoM improvement (based on the optimal
configurations found in Section~\ref{sec:opt}).  Each pair of solid
and dashed curves corresponds to a set of assumptions about $\sigmao$
and $\sigmaf$; the solid curve corresponds to fixed $\sigmaf$ and
$\rho$ in the Fisher matrix, while the dashed curve corresponds to
marginalized $\sigmaf$ (with a prior of 0.05 on $\sigmaf^2$) and
$\rho$ (with no prior).

The two black curves in the middle, labeled as (1), correspond to our
baseline assumption about the scatter: $\sigmao=0.5$ and
$\sigmaf=0.1$.  As can be seen, the uncertainty in $\rho$ barely
degrades the FoM. This minor effect is due to the small $\sigmaf$;
given such a small scatter, the follow-up mass will closely follow the
real mass even for an uncertain correlation.

On the other hand, if $\sigmaf$ is large, the follow-ups have large
intrinsic scatter and provide less constraining power.  We demonstrate
the degradation with the two blue curves at the bottom, labeled as
(2), which correspond to a high scatter in follow-ups: $\sigmaf=0.4$.
Comparing the blue solid curve to the black solid curve, we can see
that the FoM is degraded due to the larger scatter in follow-ups.  If
we further consider the uncertainty in $\rho$, the blue dashed curve
shows stronger degradation than the baseline case.

We also explore the effect of $\sigmao$.  The two red curves at the
top, labeled as (3), correspond to the case of a low scatter in
$\ln\Mobs$: $\sigmao=0.2$.  Comparing the red solid curve to the black
solid curve, we can see that the FoM is significantly improved due to
the smaller mass variance of the follow-ups. Since the follow-ups are
selected by $\Mobs$, lowering $\sigmao$ leads to follow-ups with a
less spread in mass and provides better constraints.  The red dashed
curve shows that marginalizing over $\sigmaf$ and $\rho$ only modestly
degrades this result, since in this regime both mass tracers have very
high fidelity.

We note that the FoM improvement due to a small $\sigmao$ comes from
the follow-ups rather than self-calibration.  When we lower $\sigmao$
from $0.5$ to $0.2$, the fiducial FoM from self-calibration (without
follow-ups) barely changes.  This result reflects the fact that simply
reducing the value of the scatter is not as effective as improving the
constraints on the scatter.

We also note that $\sigmaf$ and $\rho$ are assumed to be independent
of mass and redshift; detailed properties of $\sigmaf$ and $\rho$ are
beyond the scope of this work.  However, possible dependence of these
parameters on mass and redshift, if not well constrained, may severely
degrade the FoM (see, e.g., \citealt{Sahlen09}).

In summary, we have found that as long as $\sigmaf$ is sufficiently
small, the effects of $\sigmaf$ and $\rho$ are negligible, and the
uncertainty in $\rho$ has only a modest impact on the efficacy of the
follow-up observations.


\subsection{Systematic Error in Follow-up Mass Measurements}\label{sec:MassBias}

In this section, we ignore the scatter in the follow-ups and focus on
the possibility that follow-up mass measurements systematically
deviate from the true mass by a constant factor $d$. The mean of $\ln
\Mf$ takes the form
\begin{equation}\label{eq:Esys}
E[\ln \Mf|\phi(\Mobs)]=\ln d + E[\ln M |\phi(\Mobs)]. 
\end{equation}
The parameter $d$ characterizes the average systematic error of the
follow-up mass measurements and has no impact on the variance of $\ln
\Mf$. We include $\ln d$ as an additional nuisance parameter in the
Fisher matrix (Equation~\ref{eq:fisher2}) and study its impact on the
FoM. We are looking for the required constraints on $\ln d$ to avoid
severe degradation of the FoM.

When comparing Equations~\ref{eq:Mbias} and~\ref{eq:Esys}, we note
that $\ln M_{\rm bias}$ and $\ln d$ are completely degenerate in
determining $E[\ln \Mf|\phi]$. On the other hand, since $\ln d$ does
not affect the variance of $\ln\Mf$ in a bin (Equation~\ref{eq:var}),
this variance can provide information for $\ln M_{\rm bias}$ and break
the degeneracy.  Here we demonstrate again the importance of the mass
variance in follow-up observations.


\begin{figure}[t!]
\epsscale{1.2}
\plotone{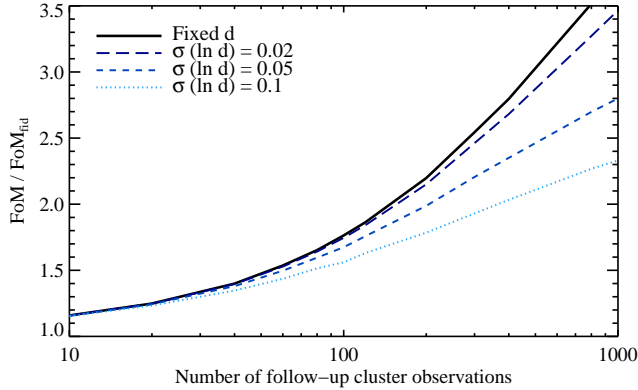}
\caption{ Impact of systematic error in follow-up mass measurements on
our optimal follow-up results.  We assume that the follow-up mass
measurements systematically deviate from the true mass by a constant
factor $d$ and compare different prior constraints on $\ln d$.
Uncertainty in $d$ can substantially degrade the efficacy of
follow-ups; for about $200$ follow-ups, a degradation of less than
$10\%$ requires $\sigma(\ln d) < 0.05$.}
\label{fig:MassBias}
\end{figure}


Figure~\ref{fig:MassBias} shows how the systematic error degrades the
efficacy of our optimal follow-up strategies. We set the fiducial
value of $\ln d$ to be 0 and compare different prior constraints on
$\ln d$. As can be seen, for approximately 200 follow-ups, a
degradation of less than $10\%$ in the FoM requires $\sigmaprior(\ln
d)<0.05$.  The required prior constraint on $\ln d$ depends on the
number of follow-up observations; larger follow-up programs require
even higher precision.  Consequently, it is very likely that the
systematic error in mass measurements will determine the efficacy of
follow-up observations.


\section{Optimizing X-ray and SZ Follow-up Programs: Observational
  Issues}
\label{sec:observability}

In previous sections, we assumed that all optically selected clusters
have an equal chance to be followed up, regardless of their mass and
redshift.  In reality, observing optically selected low-mass or
high-redshift clusters may be very difficult or even impossible with
some methods; the optimization in Section~\ref{sec:opt} is thus
impractical.  For example, observing high-redshift clusters in X-ray
requires substantially (if not prohibitively) more telescope time and
sometimes has limited improvement in parameter constraints.  Thus, we
would like to optimize the follow-up strategy considering both the FoM
and the cost of telescope time.  In this section, instead of assuming
a fixed number of follow-ups, we study how to optimize the follow-up
strategy with limited observational cost.  We first model the
observational cost in Section~\ref{sec:cost} and then demonstrate the
optimization in Section~\ref{sec:cost_opt}.

\subsection{Observability and Cost Proxies}\label{sec:cost}

For X-ray, we expect that a precise mass measurement requires certain
photon counts; therefore, we assume that the telescope time for
observing a cluster is inversely proportional to its flux of X-ray
photons.  This flux is proportional to $L_X/D_L^2$, where $L_X$ is the
X-ray luminosity and $D_L(z)$ is the luminosity distance.  We assume a
self-similar scaling relation from the fit of \cite{Vikhlinin08}, $L_X
\propto M_{500c}^{1.6} E^{1.85}(z)$.  This fit is based on the mass
with overdensity 500 times the critical density ($M_{500c}$), while we
calculate the mass function using overdensity 200 times the mean
matter density ($M_{200m}$); therefore, we convert $M_{500c}$ to
$M_{200m}$ using the fitting formula in \cite{HuKravtsov03}.  We
normalize the observational cost such that one unit corresponds to the
telescope time for observing a cluster of mass $10^{15.1}\hiMsun$ at
redshift $0.05$.  Measuring the mass of such a cluster to $10\%$
accuracy using $Y_X$ takes approximately $0.13$ ks with a single {\em
XMM} MOS camera (A. Mantz 2009, private communication)\footnote{ We
ignored the field-of-view limitations at low redshift when estimating
this time scale, which is only used for order of magnitude estimates
of the total telescope time required for a follow-up program.}.  We
note that we use $\sigmaf=0.1$ for X-ray clusters
\citep[e.g.,][]{Kravtsov06}.

For SZ, we expect that the observational time is proportional to the
inverse square of signal-to-noise ratio, S/N$\propto Y/\sqrt{\Omega}$,
where $Y$ is the integrated Compton-$y$ parameter and $\Omega$ is the
angular size of the cluster.  In virial equilibrium, $Y \propto
M^{5/3}\rho_m^{1/3}/D_A^2$, where $\rho_m(z)$ is the mean matter
density and $D_A(z)$ is the angular diameter distance.  The dependence
on angular size comes from averaging the total cluster emission over
some number of detectors. The SZ cost proxy is therefore proportional
to $D_A^2 M_{200m}^{-8/3}(1+z)^{-4}$. In addition, we exclude clusters
of redshift less than 0.1; these clusters have large angular sizes and
are contaminated by the primary CMB anisotropy. We also exclude
clusters of mass less than $10^{14.1}\hiMsun$ because they are subject
to significant background confusion \citep{Holder07}. We normalize the
observational cost such that one unit corresponds to the telescope
time for observing a cluster of mass $10^{15.1}\hiMsun$ at redshift
$0.15$.  To observe such a cluster, it takes about 30 minutes to
obtain S/N=10 with the South Pole Telescope (SPT) (D. Marrone \&
B. Benson 2009, private communication). We use a slightly larger
scatter for SZ clusters, $\sigmaf=0.2$; simulations have suggested
that this scatter may be intrinsic \citep[e.g.,][]{Shaw07}, and
projection effects can further increase the scatter.  Note that we
only consider intrinsic scatter and ignore systematic errors in the
mass measurements from both X-ray and SZ clusters.

\begin{figure*}[t!]
\epsscale{1.2}
\plotone{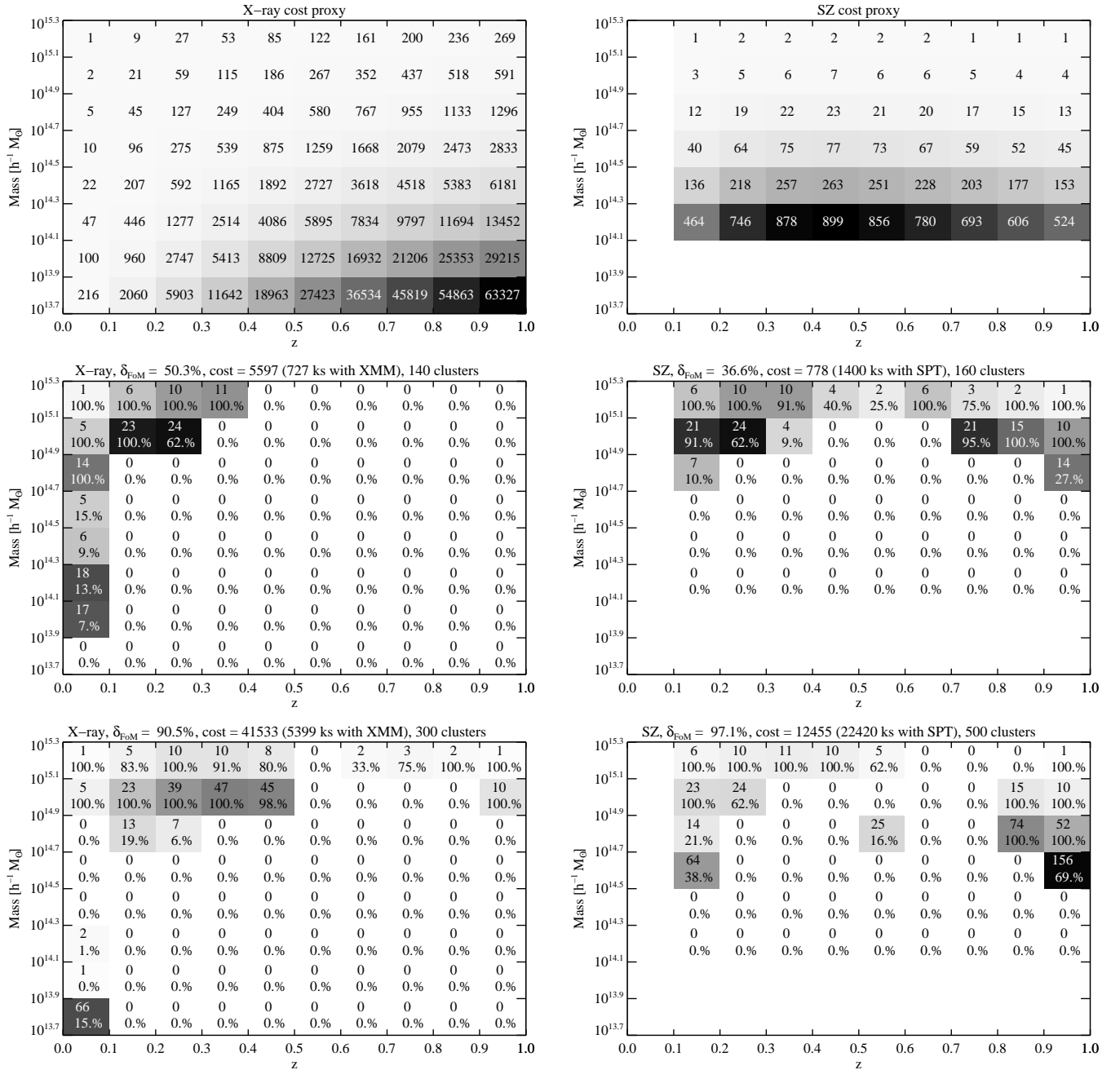}
\caption{ {\it Left:} X-ray follow-ups. {\it Right:} SZ
follow-ups. {\it Top:} observational cost proxies, which are
proportional to the required telescope time to observe a cluster and
are normalized to the lowest-redshift and highest-mass bin available
for each method.  {\it Middle} and {\it bottom}: optimal follow-up
strategies for small and large follow-up programs.  We maximize the
FoM at a given total cost and show both the number and the percentage
of follow-up targets in each bin.  As can be seen, the configurations
depend on the allowed cost.  In general, X-ray follow-up programs
favor low-redshift clusters first and extend to high-redshift as the
allowed cost increases.  On the other hand, SZ follow-ups include high
mass clusters in a wide redshift range and extend toward low mass as
the allowed cost increases.
}
\label{fig:observability}
\end{figure*}

The top panels in Figure~\ref{fig:observability} present the mass and
redshift dependence of the cost proxies.  As can be seen, X-ray and SZ
cost proxies have different patterns.  The cost of X-ray clusters
increases rapidly with redshift, while the cost of SZ clusters is
almost constant with redshift.  The latter is primarily sensitive to
mass rather than redshift; thus, for SZ, high-redshift follow-ups are
more available than low-mass ones.  We will continue to factor in the
total observational cost of a follow-up program, which is obtained by
summing over the product of cluster number and cost in each bin.


\begin{figure*}[t!]
\epsscale{1.2}
\plotone{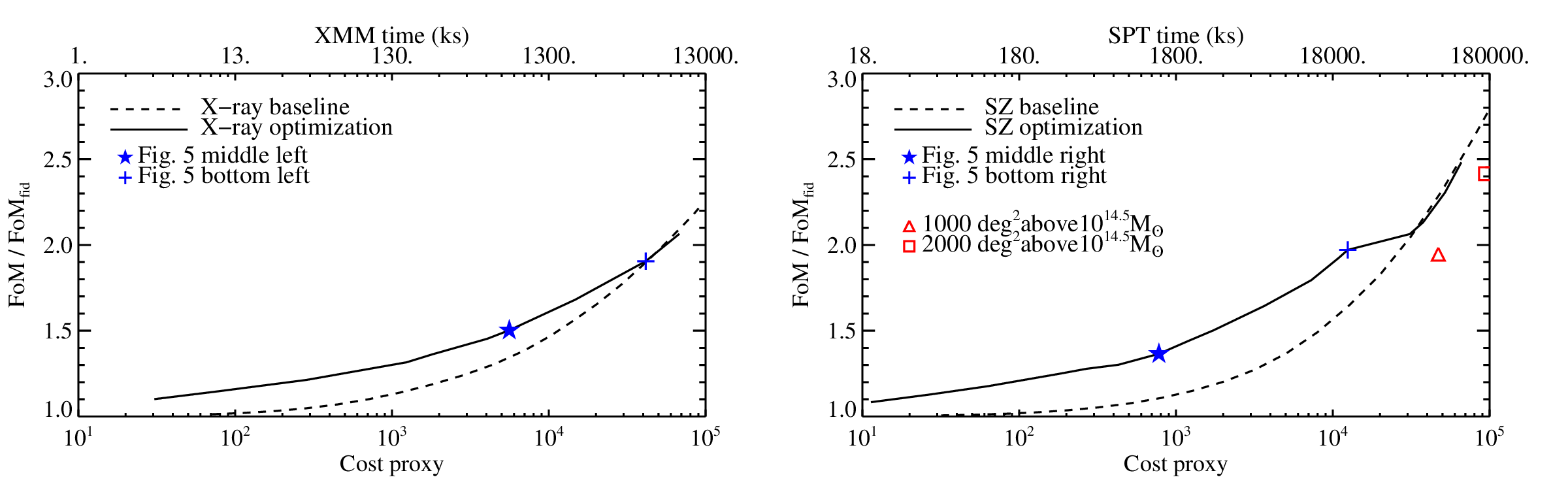}
\caption{ Improvement in the FoM as a function of observational cost.
{\it Left:} X-ray follow-ups.  {\it Right:} SZ follow-ups.  The lower
$x$-axes correspond to the cost proxies presented in
Figure~\ref{fig:observability}, while the upper $x$-axes correspond to
the observational time for two specific instruments (see the text).
The dashed curves correspond to evenly selected follow-ups with cost
below 200, while the solid curve corresponds to optimal follow-up
selections at each given cost.  Both optimized curves show changes of
slopes near the stars and crosses we marked, reflecting the changes of
follow-up strategies.  For small follow-up programs, optimal follow-up
strategies can reduce the cost by up to an order of magnitude for a
given FoM value.  On the other hand, for large follow-up programs,
optimizing with respect to cost is not essential.  In the right panel,
we also mark two points relevant to the SPT survey: 1000 $\rm deg^2$
(triangle) and 2000 $\rm deg^2$ (square), assuming a constant mass
threshold $10^{14.5}\hiMsun$.}
\label{fig:cost}
\end{figure*}

\subsection{Optimizing the Follow-up Strategy at a Given Observational Cost}\label{sec:cost_opt}

Given limited telescope time for a follow-up program, we would like to
find the strategy that maximizes the FoM. However, our optimization
algorithm in Section~\ref{sec:opt} cannot be applied directly;
sampling a configuration at a given cost is not practical, since both
the FoM and the cost depend on the configuration.  We instead use a
Monte Carlo approach: We sample a configuration and find its
corresponding point on the cost--FoM plane.  After sampling many
configurations, we can find the boundary of these points and estimate
the upper bound of the FoM at a given cost.

To generate these Monte Carlo points, we slightly modify the sampling
algorithm in Section \ref{sec:opt}.  At a given $\Nf$, we sample $\sim
10^5$ configurations and compute their corresponding cost and FoM.  We
then plot all these points on the cost--FoM plane and find the maximum
of FoM at a given cost.  To make the sampling more efficient, we
modify the algorithm to maximize (FoM/ln (cost)), which includes
moderate dependence on cost.  Since this objective function is not
well justified, we only use it in the sampling.  However, it turns out
that the configuration maximizing (FoM/ln (cost)) at a given $\Nf$
coincides with the boundary of the FoM at a given cost.  We thus
empirically propose that one can maximize (FoM/ln (cost)) at a given
$\Nf$ to design follow-ups.

The middle panels of Figure~\ref{fig:observability} show two examples
of small follow-up programs involving about 150 clusters; both are the
optimal strategies at a given cost.  We present both the number and
the percentage of follow-ups in each bin.  Comparing Figure
\ref{fig:observability} to Figure \ref{fig:opt}, we can see the impact
of observational cost on designing follow-up strategies.  For X-ray,
as expected, the high-redshift and low-mass clusters are down-weighted
because of their high cost; instead, clusters at low redshift are
chosen.  For SZ, due to its almost redshift-independent cost, massive
clusters with a wide redshift range are preferred.

The bottom panels of Figure~\ref{fig:observability} present two
examples of large follow-up programs, both of which almost double the
FoM.  As the allowed cost increases, follow-up strategies change.
These optimal configurations now extend toward the high-redshift and
low-mass corners, as in the cases in Figure \ref{fig:opt}. For X-ray,
the follow-ups include a wider range of mass and redshift.  For SZ,
less-massive clusters are included, and the configuration still favors
a wide range in redshift.

These follow-up strategies are related to our assumptions that the
observable--mass distribution depends on mass and redshift via power
laws. To constrain power laws, sampling a range of mass and redshift
is the most effective.  We emphasize that these follow-up strategies
are targeting dark energy constraints alone and do not comprehensively
consider cluster science. In reality, skipping follow-ups in some bins
may be risky for cluster science. In addition, we will need some
follow-ups in every mass and redshift bin to test our power-law
assumptions.  After the power-law assumptions are justified, we can
more confidently use our follow-up strategies to improve dark energy
constraints. We also emphasize that a single instrument is assumed for
the follow-up observations.  For SZ, however, clusters of different
redshift ranges are likely to be observed with different instruments
with different normalizations in cost.  These complications will be
instrument specific and will change the optimization.

Figure~\ref{fig:cost} shows the FoM improvement due to optimization as
a function of cost for both X-ray (left panel) and SZ (right panel)
follow-ups.  The lower $x$-axes correspond to the cost proxies
discussed in Section~\ref{sec:cost}, while the upper $x$-axes
correspond to the telescope time specific to {\em XMM} and SPT.  We
compare the optimal cases (solid curves) with the baseline cases
(dashed curves).  The baseline cases correspond to equal number of
follow-ups in all mass and redshift bins with cost less than 200. The
importance of optimizing follow-up strategies is abundantly clear: to
achieve a specified FoM, our optimal strategies can reduce the
required telescope time by about an order of magnitude for small
follow-up programs.

In Figure~\ref{fig:cost}, the optimal cases for X-ray and SZ show
different features.  The stars and crosses mark the examples we have
shown in Figure~\ref{fig:observability}.  As can be seen, for X-ray,
as the allowed cost increases, the FoM increases less rapidly than SZ.
Both curves show slope changes, which are caused by the changes in
configurations, as we discuss below.

For X-ray, when the allowed cost is below $10^4$, the most effective
strategy is to tighten the constraints on the observable--mass
distribution in low-redshift bins (as shown in the middle left panel
in Figure~\ref{fig:observability}).  When the allowed cost is high
enough, the follow-ups can afford to constrain both low and high
redshift bins (as shown in the bottom left panel).  Since constraining
two extreme redshift regimes gives much better constraints on the
evolution, the slope of the FoM increases.  Nevertheless, this slope
increase is very close to the point where the optimal case approaches
the baseline case.  With such a large allocation of telescope time,
optimization is no longer essential, and the uniform sampling can
achieve the same FoM.

For SZ, two obvious slope changes can be seen.  The first one occurs
near the cost of 1000.  Below this cost, only massive clusters are
chosen (as shown in the middle right panel of
Figure~\ref{fig:observability}).  Above this cost, less-massive
clusters become affordable (as shown in the bottom right panel), and
the mass dependence of the observable--mass distribution is better
constrained, leading to the slope increase of the FoM.  The second
change occurs near the cost of $10^4$, where the slope suddenly drops
and the optimal case approaches the baseline case.  At this point, we
exhaust the information from the two redshift ends, and sampling the
middle regime cannot make significant improvement.  The lack of
further improvement leads to the decrease of slope, and optimization
is no longer essential when this amount of telescope time is
available.

Comparing X-ray and SZ, we can view the design of cost-effective
follow-up strategies as a trade-off between constraining mass
dependence and constraining redshift evolution of the observable--mass
distribution.  When the cost is more sensitive to redshift than mass,
as in the case of X-ray, one should prioritize the constraints on mass
dependence regarding limited telescope time.  On the other hand, when
the cost is more sensitive to mass than redshift, as in the case of
SZ, redshift evolution should be prioritized.

In the right panel of Figure~\ref{fig:cost}, we add two points as
references: the red triangle and the red square present the cases,
respectively, of 1000 and 2000 ${\rm deg^2}$ of SZ follow-ups; these
assumptions are relevant for SPT.  We assume a mass threshold of
$10^{14.5}\hiMsun$, which roughly corresponds to the mass threshold of
the ongoing SPT survey (L. Shaw \& B. Benson 2009, private
communication). However, here we assume that the follow-up mass
measurements have a constant scatter and no systematic error.  In
reality, the SPT survey may not have the same precision in mass
measurements for all clusters, and degradation due to inaccurate mass
estimates is possible.

\begin{figure}[t!]
\epsscale{1.2}
\plotone{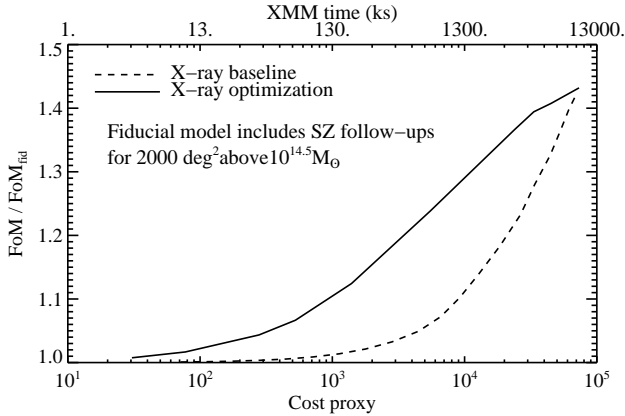}
\caption{ Improvement in the FoM as a function of observational cost
from X-ray follow-ups in addition to $2000 \deg^2$ SZ observations
above $10^{14.5}\hiMsun$.  These follow-up strategies all focus on
low-redshift and low-mass clusters in order to provide complementary
information to SZ follow-ups.  We note that in this regime, the
improvement in the FoM is less significant than previous cases because
the fiducial FoM already includes substantial information from
follow-ups.  }
\label{fig:SPT_X}
\end{figure}

Finally, we study the complementarity between X-ray and SZ follow-ups.
We assume that $2000 \deg^2$ of the survey field is followed up by SZ
with a mass threshold of $10^{14.5}\hiMsun$.  We then optimize the
X-ray follow-ups for clusters that are not observed by SZ.  In Figure
\ref{fig:SPT_X}, we show the FoM improvement due to these additional
X-ray follow-ups.  The fiducial FoM corresponds to a DES-like survey
with $2000 \deg^2$ SZ follow-ups, and the baseline case corresponds to
evenly selected targets with cost below 200.  To provide complementary
information to SZ follow-ups, the optimal X-ray follow-up strategies
would focus on low-redshift and low-mass clusters.  We note that since
the fiducial FoM already includes the information from SZ follow-ups,
the improvement in the FoM is less significant than that in Figure
\ref{fig:cost}.

In this calculation, we assume the SZ follow-up mass measurements have
20\% scatter and no systematic error.  This is an optimistic
assumption for SPT; in reality, clusters observed by SPT may have
larger scatter and more complicated sources of systematic error.
Therefore, our calculation only represents a limiting case.  On the
other hand, our calculation assumes that X-ray and SZ follow-ups are
not overlapping.  In reality, some clusters will be observed with both
methods.  In principle, one can calculate the combined information
from both X-ray and SZ follow-ups and optimize follow-up strategies by
including these doubly followed-up clusters.  However, since we assume
very accurate mass inference for SZ clusters, X-ray observations will
add little to it.  In addition, computing the information from both
follow-up methods will require modeling the full covariance between
optical, X-ray, and SZ observables, which is beyond the scope of the
current work.

\section{Discussion}\label{sec:discussion}

In Section~\ref{sec:correlation}, we studied the impact of the
correlation between different mass proxies.  In principle, since the
true mass is not observable, $\rho$ cannot be directly measured.
However, this correlation can be studied using consistency of scaling
relations for different mass proxies.  For example, \cite{Rykoff08b}
studied the scaling relation of the mean X-ray luminosity and mean
weak lensing mass for optically selected samples, which are binned by
optical richness.  They compared this scaling relation with the one
derived from X-ray selected samples and found that the correlation
between X-ray luminosity and optical richness is consistent with zero.
Although these authors did not provide constraints on the correlation,
they demonstrated an effective way to study it.  In addition,
\cite{Rozo08a} used a similar analysis to obtain the constraints on
the correlation between X-ray luminosity and mass for a given optical
richness, finding $\rho_{L_X, M|N} \ge 0.85$.

This correlation can also be studied with simulations.  For example,
the results of \cite{Wechsler06} imply a slight anti-correlation: at a
given halo mass, halos with high concentration have lower richness on
average.  \cite{CohnWhite09} studied the joint SZ and optical cluster
finding in simulations.  They demonstrated that the cluster mass
estimates from optical richness and SZ flux are positively correlated.
Detailed comparisons for different mass proxies, however, will require
further exploration \citep[e.g.,][]{Stanek09}.

In Section~\ref{sec:MassBias}, we studied the impact of the systematic
error of follow-up mass measurements.  Different follow-up methods
have different sources of systematic error; here we compare several
different mass proxies studied in the literature. \cite{Nagai07}
simulated the X-ray mass measurements and found that the total cluster
mass derived from hydrostatic equilibrium is systematically lower than
the true mass by about $5\%$--$20\%$ \citep[also see,
e.g.,][]{Rasia06,Mahdavi08}.  They also found that the deviation is
less significant for relaxed systems and for the inner regions of the
clusters. This underestimate can be attributed to the non-hydrostatic
state of the intracluster medium that provides additional pressure
support \citep[e.g.,][]{Evrard90,Lau09}.  On the other hand, they
found that the estimate of the mass of the intracluster medium
($M_{\rm gas}$) is robust.

\cite{Vikhlinin08} used multiple X-ray indicators, including $M_{\rm
gas}$, the temperature $T_X$, and estimated total thermal energy $Y_X
= M_{\rm gas} \times T_X$ to calibrate the cluster mass. These mass
indicators have been shown by simulations to have a tight scaling
relation with the total mass \citep[e.g.,][]{Kravtsov06}.
\cite{Vikhlinin08} also calibrated the total mass with low-redshift
samples and cross-checked it with the weak lensing results.
Therefore, X-ray clusters, when carefully calibrated, are likely to
provide the most robust mass proxy and the most ideal method for
follow-up observations.

On the other hand, SZ observations are still limited by statistics and
have few observational studies on the scaling relation and the cluster
profile \citep[e.g.,][]{Mroczkowski08,Bonamente08}.  Their utility is
thus yet to be fully demonstrated.  In addition, \cite{RuddNagai09}
simulated the two-temperature model for clusters found that the
non-equipartition of electrons and ions may lead to $5\%$
underestimate of the mass derived from SZ.  Nevertheless, simulations
have shown that the scatter of the SZ mass proxy is small. If the
systematic error can be well constrained, SZ follow-ups may become
very influential given their statistical power in the near future.

Another possibility of follow-up observations is weak lensing mass
measurements of individual clusters. \cite{Hoekstra07} and
\cite{Zhang08} compared the mass measurements from weak lensing and
X-ray, finding good agreement.  However, weak lensing mass
measurements usually have $20\%$ uncertainties due to projection along
the line of sight, and the current statistics are still low. In
addition, detailed understanding of the photometric redshift
properties of source galaxies is required to avoid systematic error in
the recovered weak lensing mass \citep[see e.g.,][]{Mandelbaum08}.

Stacked weak lensing analysis has been used to measure the mean mass
of the clusters for a given optical richness
\citep[e.g.,][]{Johnston07,Mandelbaum08b}.  This method does not
suffer from projection effects caused by uncorrelated structure and
allows one to estimate the mass of low-mass clusters, for which
individual weak lensing cannot be detected.  However, the stacked
analysis cannot provide the variance of mass, which, as we have shown,
contains important information.  If the variance of mass can be
determined using this method in the future (for example, by
resampling), the constraining power of this method will be improved.

Finally, we note that one caveat of our results is the assumption that
the optical richness--mass distribution is well described by a
log-normal distribution and the scaling and scatter follow power laws.
The validity of these assumptions will need to be tested explicitly
with both simulations and observations.


\section{Conclusions}\label{sec:summary}

We studied the impacts of follow-up observations---more precise
measurements of cluster mass---on the constraining power of large
optical cluster surveys.  Considering the self-calibrated cluster
abundance data from the DES, we demonstrated that the dark energy FoM
can be significantly improved.  Our primary findings are

\begin{enumerate}

\item Optimal target-selection strategies are essential for maximizing
the power of modestly sized follow-up programs.  For instance, $100$
optimally selected follow-ups can improve the FoM of a DES-like survey
by up to $76.4\%$, which is compared to a $40.3\%$ improvement due to
evenly selected follow-ups.  Random sub-sampling of the cluster
catalog is even less effective.  Generally speaking, one should always
follow up low-redshift clusters first, and then extend to the
higher-redshift and lower-mass regime (Sections~\ref{sec:follow-up}
and \ref{sec:opt}).

\item The scatter of the follow-up mass proxy and the covariance
between the optical richness and the follow-up mass proxy have only
modest effects on the FoM, provided that the follow-up mass proxy has
sufficiently small scatter.  On the other hand, although lowering the
scatter of optical richness does not change the baseline
self-calibration results, it will significantly enhance the efficacy
of follow-ups (Section~\ref{sec:correlation}).

\item Systematic error in follow-up mass measurements should be
controlled at the $5\%$ level to avoid severe degradation.  In
addition, if only the mean of cluster mass is measured, the systematic
error of follow-up mass proxy will be degenerate with the systematic
error of optical richness; measuring the variance of cluster mass can
break this degeneracy (Section~\ref{sec:MassBias}).

\item We explored observational issues to propose more practical X-ray
and SZ follow-up programs.  The observational costs of X-ray and SZ
are, respectively, sensitive to redshift and mass, which in turn leads
to different follow-up strategies.  To achieve $50\%$ improvement in
the FoM, the most cost-effective follow-up strategy involves
approximately 200 low-redshift X-ray clusters or massive SZ clusters.
In general, our optimal strategies can reduce the observational cost
required to achieve given dark energy constraints by up to an order of
magnitude (Section~\ref{sec:observability}).

\end{enumerate}

A follow-up mass tracer that has systematic error understood at the
$5\%$ level will substantially benefit optical cluster surveys.  On
the other hand, reducing the scatter of optical mass tracer will
significantly improve the efficacy of optically selected follow-ups.
Current observational resources allow a few hundred low-redshift X-ray
clusters, and in the near future hundreds or thousands of SZ clusters
will become available.  Therefore, detailed follow-up studies of a
small but optimally-selected cluster sub-sample have the potential to
be a powerful complement to current and imminent cluster surveys.

\acknowledgments We thank Adam Mantz, Evan Million, Dan Marrone, and
Brad Benson for help with estimating the observational cost of X-ray
and SZ follow-ups, and Eli Rykoff for help with IDL routines.  We also
thank Michael Busha, Gus Evrard, Dragan Huterer, David Rapetti, and
Laurie Shaw for helpful suggestions.  We are grateful to the anonymous
referee for helpful comments.  We are also grateful for the support
and hospitality of both CCAPP and KIPAC during collaboration visits.
H.W. and R.H.W. were supported in part by the U.S. Department of
Energy under contract number DE-AC02-76SF00515.  H.W. received
additional support from McMicking and Gabilan Stanford Graduate
Fellowships.  R.H.W. received support from a Terman Fellowship from
Stanford University. E.R. is supported by the Center for Cosmology and
Astroparticle Physics at the Ohio State University and by NSF grant
AST 0707985.

\bibliography{ms}
\end{document}